# Discovering New Sentiments from the Social Web [1]

JOAQUÍN BORREGO DÍAZ, University of Seville
JUAN GALÁN PÁEZ, University of Seville

## 1. INTRODUCTION

A persistent challenge in Complex Systems (CS) research is the phenomenological reconstruction of systems from raw data. In order to face the problem, the use of sound features to reason on the system from data processing is a key step. In the specific case of complex societal systems, sentiment analysis allows to mirror (part of) the affective dimension. However it is not reasonable to think that individual sentiment categorization can encompass the new affective phenomena in digital social networks.

The present papers addresses the problem of isolating sentiment concepts which emerge in social networks. In an analogy to Artificial Intelligent Singularity, we propose the study and analysis of these new complex sentiment structures and how they are similar to or diverge from classic conceptual structures associated to sentiment lexicons. The conjecture is that it is highly probable that hypercomplex sentiment structures -not explained with human categorizations- emerge from high dynamic social information networks. Roughly speaking, new sentiment can emerge from the new global nervous systems [Pentland 2012] as it occurs in humans.

### 1.1 Motivation of the work

The rise of impressive social media platforms as Twitter increases the interest in the development of sentiment analysis technologies and tools in order to understand how certain social digital events occur. Although sentiment is usually opposed to knowledge, semantic structure of human sentiments can be considered as a Knowledge organization about them. Formal Concept Analysis (FCA) [Ganter and Wille 1999] can used to organise knowledge and extract new concepts from data. If social messages are considered from the point of view of sentiment, these messages may be considered as examples of knowledge with sentiment features. On the one hand, opinion lexicons (bag of tagged words) will data source to apply FCA tools. On the other hand, structured English lexical database as WordNet provide a robust semantic structure to conceptualize old (and new) kind of sentiment concepts. Our motivation is to study if FCA provides a nice framework to address these questions

By means of the intensive use of Formal Concept Analysis [Ganter and Wille 1999], which provides mathematical tools for detecting newly minted sentiments, established categorizations for sentiment are compared. Once selected the lexicon, we aim to estimate how such new concept structures for digital societies converge to (or diverge from) single-agent based ones. Experiments to discover new sentiment structures about specific events from information dynamics in Twitter are shown (see in Fig. 3 a snapshot of conceptual sentiment structure for Syria conflict).

## 2. FORMAL CONCEPT ANALYSIS

FCA mathematizes the philosophical understanding of a concept as a unit of thoughts composed of two parts: the extent and the intent. The extent covers all objects belonging to the concept, while the intent comprises all common attributes valid for all the objects under consideration [Ganter and Wille 1999]. It also allows the computation of concept hierarchies from data tables. It represents an automated conceptual learning theory, which allows to detect and describes regularities and structures

| | River | Coast | Sea |
|---|---|---|---|
| Carp | X | | |
| Escatofagus | X | X | |
| Bream | | X | X |
| Sparus | | X | X |
| eel | X | X | X |

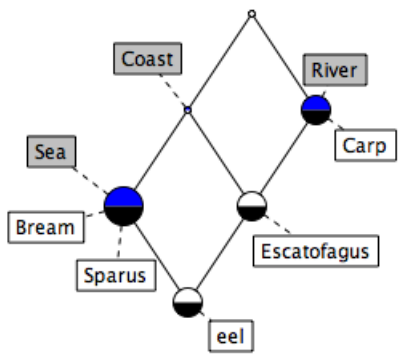

Fig. 1. Formal context of fish, and its concept lattice

of concepts. Also, it provides knowledge bases (sets of logical implications between attributes) as Stem Basis, Luxenburger Basis. The bass data structures in FCA are the Formal Context $(O, A, I)$ and The Concept Lattice

A *formal context* $M = (O, A, I)$ consists of two sets, $O$ (objects) and $A$ (attributes), and a relation $I \subseteq O \times A$. Finite contexts can be represented by a 1-0-table (identifying $I$ with a boolean function on $O \times A$). Given $X \subseteq O$ and $Y \subseteq A$, it defines

$$X' = \{a \in A \mid oIa \text{ for all } o \in X\} \text{ and } Y' = \{o \in O \mid oIa \text{ for all } a \in Y\}$$

The main goal of FCA is the computation of the concept lattice associated with the context. A (formal) concept is a pair $(X, Y)$ such that $X' = Y$ and $Y' = X$. For example, the concept lattice from the formal context of fishes of Fig. 1, left (attributes are understood as ”live in”) is depicted in Fig. 1, right. Each node is a concept, and its intension (or extension) can be formed by the set of attributes (or objects) included along the path to the top (or bottom). For example, the bottom concept $(\{eel\}, \{Coast, Sea, River\})$ is the concept *euryhaline fish*. CL contains every concept that can be extracted from the context. As well, concepts are defined but it is possible that no specific term (word) exists to denote it.

The concept lattice represents a semantic organization of the (unstructured) information source, by providing a partially ordered structure of concepts (a hierarchy of concepts). From this point of view, it can be considered a complex semantic network, a network which presents different topology depending on the nature of the data [Aranda-Corral et al. 2013a]. However, a question leads from this approach when a big amount of information about the CS is provided.

### 2.1 Concept Lattice versus Knowledge Robustness

On the one hand, to achieve a sound (qualitative) phenomenological reconstruction of the CS requires of a a sound selection of the features (qualitative) to be computed/studied. Therefore, some questions arise:

—How is it decided if the feature selection is sound?

—How is the soundness of the qualitative representation/reasoning model obtained from a feature selection analyzed?

—Specifically: Do sound qualitative modelizations share similar structure/properties?

There exists a great amount of knowledge on existing real semantic networks and how the topology of these networks is deeply related with the soundness of the representation. In our case, semantic networks are synthesized from raw and selected data, so a kind of reverse knowledge engineering has to be made. That is, it seem that the concept lattice have to share topological properties with others successful semantic network based representations. In [Aranda-Corral et al. 2013a] the authors study the called *Scale-Free Conceptualization Hypothesis (SFXH)*: Only if the attribute set selected to observe the Complex System is computable, objective, and induces a Concept Lattice that provides a sound analysis of the CS (from the point of view of some type of BR), then its degree-distribution is scale-free. [Aranda-Corral et al. 2013a].

The issues involved in the use of SFCH as primary test to estimate the fit of semantic representation mainly are:

—SFCH works as test of information quality

—Monster model (that is, the concept lattice built from all data without previous selection of relevant features) has to be useful for capturing CS dynamics

## 3. A FORMAL CONTEXT FOR WORDNET

It is possible to apply FCA in order to study our language and relate formal concepts with those of a real language. With this purpose the lexical database of English *WordNet* was considered. WordNet basically structures the language in form of words (*Lemmas*) grouped in sets of cognitive synonyms (*Synsets*).

A huge formal context was built by considering lemmas as objects and synsets as attributes and the associated concept lattice was computed. It is worthy to note that the lattice shows a strong relationship between formal concepts and language concepts. Also this concept lattice achieves the aforementioned SFCH, that is, its node's degree distribution follows a power lay distribution. Therefore we can claim that, as expected, WordNet presents well structure semantics and can be a useful tool in order to give some structure to opinion analysis tasks.

## 4. FORMAL CONCEPT ANALYSIS (FCA) FOR SENTIMENT DISCOVERING

In this section we succinctly describe the experiments carried out with the aim of enhancing opinion lexicons by providing them with a semantic structure by means of FCA and WordNet.

### 4.1 Enhancing Opinion Lexicon

Opinion lexicons basically consist on lists of words (relevant for sentiment detection) that are annotated with their polarity (how positive or negative they are). As it was mentioned, WordNet provides semantic relations between words, and the concept lattice provides structure. In this way it is possible to enhance an opinion lexicon by means of these tools.

### 4.2 Applying FCA: Sentiment Lattices

A *Sentiment Lattice* is a subset of the concept lattice associated to WordNet. This subset is obtained by considering only the concepts related with the words of the opinion lexicon. The concepts of this lattice are called *sentiment concepts*, and each of them has a polarity associated. The valuation of each concept comes from the aggregation of the sentiment of each word existing in the extension of the concept.

Some experiments with three different opinion lexicon where performed in order to show the usefulness of FCA as tool for testing the consistency of sentiment lexicons as well as for analyzing information item's sentiment. Information items considered for testing are tweets due to the fact that are short concentrated pieces of information. As the pieces of text are quite short, no natural language processing (NLP) approaches (as identifying parts of speech, etc.) have been considered. Thus in order to collect pieces of sentiment within tweets, the approach has been to look for occurrences of isolated sentiment words in the information items. After that, the sentiment word set associated to each information piece (tweet) is allocated within the sentiment lattice in the corresponding concept.

In Fig. 2 the results of the analysis of the three different opinion lexicon are shown. In order to test the usefulness of these lexicons to be used on twitter, some measures has been considered:

| Sentiment lexicon | Words | Tagged tweets | Ambiguous sentiment concept | Specific sentiment concept | New sentiment concept |
|---|---|---|---|---|---|
| Bing Liu's Opinion Lexicon | 6,800 | 2,950 | 800 | 2,000 | 150 |
| SentiWordNet | 39,000 | 6,350 | 5,000 | 900 | 450 |
| AFINN-111 Lexicon | 2,500 | 3,900 | 1,000 | 2,800 | 100 |

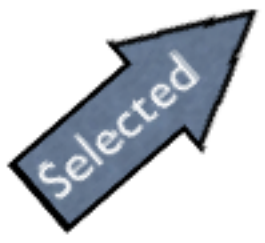

*Bing Liu's Opinion Lexicon:* http://www.cs.uic.edu/~liub/FBS/sentiment-analysis.html
*SentiWordNet:* http://sentiwordnet.isti.cnr.it/
*AFINN-111:* http://www2.imm.dtu.dk/pubdb/views/publication_details.php?id=6010

Fig. 2. Data on different opinion lexicon and their performance allocating sentiment concepts to tweets (11,500 tweets aprox.)

—Tagged tweets: Polarised tweets, that is, tweets containing some sentiment word. The rest are not polarised, thus they are not considered.

—Specific sentiment concept: Tweets allocated in an existent concept within the lattice.

—New sentiment concept: Tweets whose sentiment words produce a new concept but respecting the existing structure.

—Ambiguous sentiment concept: Tweets which would be allocated in different parts of the lattice or can not be allocated.

After this analysis we can conclude that for our purposes, the most interesting sentiment lexicon is *AFINN-111* (this does not mean is the best). Although SentiWordNet seems to be the most promising it showed to polarize too many words, producing some ambiguity.

### 4.3 A case Study

After selecting a proper sentiment lexicon, an experiment was performed with a tweet set of around $15,000$ English tweets, containing the topic ”Syria” collected in 05/20/2013 during $6$ hours. Finally, the whole tweet set is allocated on the sentiment lattice in order to populate it. The graph shown in figure 3 is a weighted sentiment lattice, where the weight (node size) shows the number of tweets allocated in each concept and the color indicates the polarity of the concept (green is positive and ted negative). Once this populated sentiment lattice is obtained, it is possible to study the semantic relationship between tweets allocated in neighboring concepts as well as the new concepts generated which can give insights of new word usage that have not been considered previously.

Authors have designed FCA-based tools and methods to study CS from agent interaction [Aranda-Corral et al. 2013c] to global semantic scale [Aranda-Corral et al. 2013a], as well as to successfully simulate the predictive human behavior [Aranda-Corral et al. 2013b].

## 5. CONCLUSIONS

The application of FCA tools to Sentiment Analysis provides structures for Knowledge Organization that can be analyzed by means of both network and computational logic tools. Also, sentiment lattices provide a partially ordered structure on feeling exposed in information items. This structure allows us to discover semantic relationships among information items that are tightly related with sentiment. As discussed above, even new sentiment concepts emerge and sentiment lattice show them.

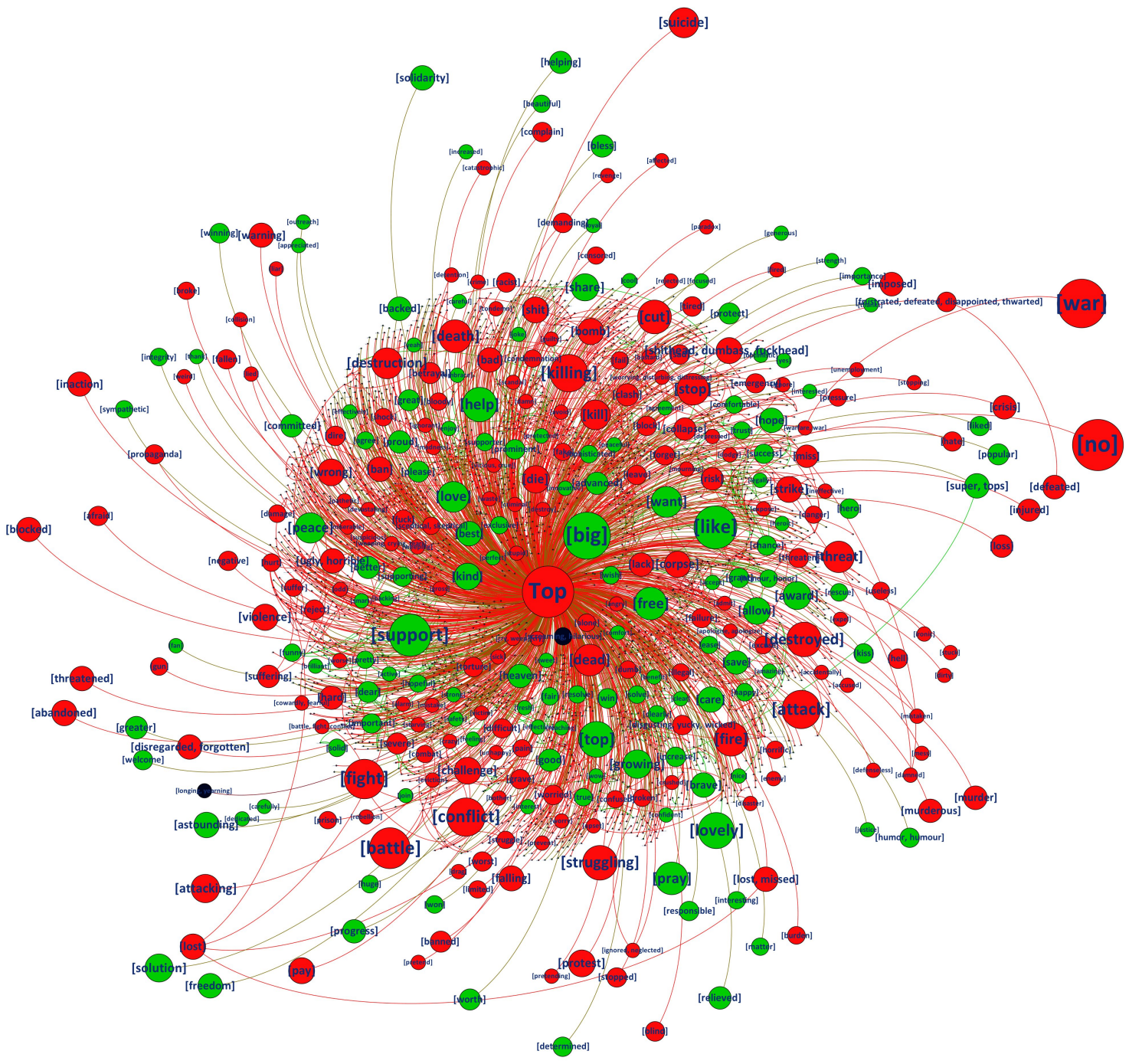

Fig. 3. A snapshot of Sentiment Structure about Syria conflict in Twitter

Interesting insights can be obtained by analyzing how the sentiment graph evolves over time. Using network-based metrics can be explored how sentiment lattices evolves while the event *live* on the social stream. Using computational logic it is possible to conjecture how sentiment concepts evolves, by extracting and analyzing the sentiment lattice trend (with knowledge basis as Stem Basis).

With respect to the emergence of new kind of sentiment concepts, the semantic analysis of these in ICT social systems is made by means of FCA. In fact, FCA allows extracting semantic sentiment concepts from noisy information items. The new elements are also semantically related with information items and among them. Future work pass through the use NLP tools to enhance the useful information retrieved from an information item, and, lastly, to use Bounded Rationality techniques to forecast the evolution of a sentiment concept within an ICT social system (as in [Aranda-Corral et al. 2013b]).

## REFERENCES

G. Aranda-Corral, J. Borrego-Díaz, and J. Galán-Páez. 2013a. On the Phenomenological Reconstruction of Complex Systems—The Scale-Free Conceptualization Hypothesis. *Systems Research and Behavioral Science* 30, 6 (2013), 716–734. DOI:http://dx.doi.org/10.1002/sres.2240

G. Aranda-Corral, J. Borrego-Díaz, and J. Galán-Páez. 2013b. Complex Concept Lattices for Simulating Human Prediction in Sport. *J. Syst. Science and Complexity* 26, 1 (2013), 117–136.

G. Aranda-Corral, J. Borrego-Díaz, and J. Giráldez-Cru. 2013c. Agent-mediated shared conceptualizations in tagging services. *Multimedia Tools and Applications* 65, 1 (2013), 5–28. DOI:http://dx.doi.org/10.1007/s11042-012-1146-5

B. Ganter and R. Wille. 1999. *Formal Concept Analysis: Mathematical Foundations*. Springer, Berlin/Heidelberg.

A. Pentland. 2012. Society's Nervous System: Building Effective Government, Energy, and Public Health Systems. *Computer* 45, 1 (2012), 31–38. DOI:http://dx.doi.org/10.1109/MC.2011.299